\documentclass[twocolumn,times]{aastex7}
\usepackage{caption}
\usepackage{color}
\usepackage{soul}

\begin{document}

\title{Spatially and Dynamically Extended Molecular Gas in Stephan’s Quintet Revealed by ALMA CO(1–0) Total Power Mapping}

\author[orcid=0000-0002-8868-1255]{Fumiya Maeda}
\affiliation{Research Center for Physics and Mathematics, Osaka Electro-Communication University, 18-8 Hatsucho, Neyagawa, 572-8530, Osaka, Japan}
\email[show]{fmaeda@osakac.ac.jp}  

\author[orcid=0009-0007-2493-0973]{Shinya Komugi}
\affiliation{Division of Liberal Arts, Kogakuin University, 2665-1 Nakano-cho, Hachioji, Tokyo 192-0015, Japan}
\email{skomugi@cc.kogakuin.ac.jp}

\author[orcid=0000-0002-3373-6538]{Kazuyuki Muraoka}
\affiliation{Department of Physics, Graduate School of Science, Osaka Metropolitan University, 3-3-138 Sugimoto, Sumiyoshi-ku, Osaka, Osaka 558-8585, Japan}
\email{kmuraoka@omu.ac.jp}  

\author{Misaki Yamamoto}
\affiliation{Department of Physics, Graduate School of Science, Osaka Metropolitan University, 3-3-138 Sugimoto, Sumiyoshi-ku, Osaka, Osaka 558-8585, Japan}
\email{}  

\author[orcid=0000-0002-1639-1515]{Fumi Egusa}
\affiliation{Institute of Astronomy, Graduate School of Science, The University of Tokyo, 2-21-1 Osawa, Mitaka, Tokyo 181-0015, Japan}
\email{} 

\author[0000-0003-3844-1517]{Kouji Ohta}
\affiliation{Department of Astronomy, Kyoto University, Kitashirakawa-Oiwake-Cho, Sakyo-ku, Kyoto 606-8502, Japan}
\email{} 

\author[orcid=0000-0003-3983-5438,sname=Asada,gname=Yoshihisa]{Yoshihisa Asada}
\affiliation{Waseda Research Institute for Science and Engineering, Faculty of Science and Engineering, Waseda University,\\ 3-4-1 Okubo, Shinjuku, Tokyo 169-8555, Japan}
\email{asada@kusastro.kyoto-u.ac.jp}  

\author[orcid=0009-0005-2452-6183]{Asao Habe}
\affiliation{Faculty of Science, Hokkaido University, Kita 10 Nishi 9, Kitaku, Sapporo, 060-0810, Sapporo, Japan}
\email{}

\author[0000-0001-6469-8725]{Bunyo Hatsukade}
\affiliation{National Astronomical Observatory of Japan, 2-21-1 Osawa, Mitaka, Tokyo 181-8588, Japan}
\affiliation{Graduate Institute for Advanced Studies, SOKENDAI, Osawa, Mitaka, Tokyo 181-8588, Japan}
\affiliation{Department of Astronomy, Graduate School of Science, The University of Tokyo, 7-3-1 Hongo, Bunkyo-ku, Tokyo 133-0033, Japan}
\email{bunyo.hatsukade@nao.ac.jp}

\author[0000-0002-2699-4862]{Hiroyuki Kaneko}
\affiliation{Collage of Creative Studies, Niigata University, 8050
Ikarashi 2-no-cho, Nishi-ku, Niigata, 950-2181, Japan}
\email{kaneko.hiroyuki.astro@gmail.com}

\author[orcid=0000-0003-3990-1204]{Masato I.N. Kobayashi}
\affiliation{I. Physikalisches Institut, Universität zu Köln, Zülpicher Str. 77, D-50937 Köln, Germany}
\email{kobayashi@ph1.uni-koeln.de} 

\author[orcid=0000-0002-4052-2394]{Kotaro Kohno}
\affiliation{Institute of Astronomy, Graduate School of Science, The University of Tokyo, 2-21-1 Osawa, Mitaka, Tokyo 181-0015, Japan}
\affiliation{Research Center for the Early Universe, Graduate School of Science, The University of Tokyo, 7-3-1 Hongo, Bunkyo-ku, Tokyo 113-0033, Japan}\email{kkohno@ioa.s.u-tokyo.ac.jp}

\author[orcid=0000-0002-4098-8100]{Ayu Konishi}
\affiliation{Department of Physics, Graduate School of Science, Osaka Metropolitan University, 3-3-138 Sugimoto, Sumiyoshi-ku, Osaka, Osaka 558-8585, Japan}
\email{sw23227n@st.omu.ac.jp}

\author[orcid=0000-0001-5509-8218]{Ren Matsusaka}
\affiliation{Institute of Astronomy, Graduate School of Science, The University of Tokyo, 2-21-1 Osawa, Mitaka, Tokyo 181-0015, Japan}
\email{}

\author[orcid=0000-0003-3932-0952]{Kana Morokuma-Matsui}
\affiliation{Institute of Astronomy, Graduate School of Science, The University of Tokyo, 2-21-1 Osawa, Mitaka, Tokyo 181-0015, Japan}
\email{kanamoro@ioa.s.u-tokyo.ac.jp}

\author[orcid=0000-0002-6939-0372]{Kouichiro Nakanishi}
\affiliation{National Astronomical Observatory of Japan, 2-21-1 Osawa,
Mitaka, Tokyo 181-8588, Japan}
\affiliation{Graduate Institute for Advanced Studies, SOKENDAI, Osawa, Mitaka, Tokyo 181-8588, Japan}
\email{}

\author[orcid=0000-0001-9016-2641]{Tomoka Tosaki}
\affiliation{Department of Geoscience, Joetsu University of Education, Yamayashiki-machi, Joetsu, Niigata 943-8512, Japan}
\email{tosaki@juen.ac.jp}

\author[0000-0002-0498-5041]{Akiyoshi Tsujita}
\affiliation{Institute of Astronomy, Graduate School of Science, The University of Tokyo, 2-21-1 Osawa, Mitaka, Tokyo 181-0015, Japan}
\email{tsujita@ioa.s.u-tokyo.ac.jp}

\begin{abstract}

We present ALMA Total Power CO(1–0) mapping of Stephan’s Quintet (SQ), a prototypical compact galaxy group, with a uniform noise level at a spatial scale of $\sim$25~kpc. 
These observations provide the first complete view of molecular gas across the whole system. Molecular gas is found to spread over a wide area ($\sim120\times80$~kpc), mainly over the two main member galaxies (NGC~7318B and 7319), but also in the shocked ridges between these galaxies, the tidal tail, and also in intergalactic regions north of the tail.
The total CO(1–0) luminosity is $(2.47\pm0.12)\times10^9~\mathrm{K~km~s^{-1}~pc^2}$, corresponding to a molecular gas mass of $(1.07\pm0.05)\times10^{10}~M_\odot$ assuming the Galactic CO-to-H$_2$ conversion factor. 
The global star formation efficiency of SQ is estimated at $0.29$–$0.70~\mathrm{Gyr^{-1}}$, comparable to or lower than that of nearby star-forming galaxies. Molecular gas spans a velocity range of $\sim1300~\mathrm{km~s^{-1}}$, which can be divided into three components (low, mid, high).
The low- and mid-velocity components, linked to NGC~7318B and the ridge, show relatively active star formation, whereas the high-velocity component, associated with NGC~7319, shows suppressed star formation. Our mapping reveals molecular gas extending $\sim$100~kpc in projection along the inner tail and north of it, containing $(1.64\pm0.08)\times10^9~M_\odot$ (15\% of total) with low velocity dispersion ($\sim20~\mathrm{km~s^{-1}}$) and ongoing star formation. 
While previous studies suggested in situ molecular gas formation in the tail, our data suggest an additional contribution from gas stripped from NGC~7319.

\end{abstract}

\keywords{
}


\section{Introduction} \label{sec: intro}

In the early universe, galaxies were closely packed, and frequent interactions and mergers are thought to have driven the growth of massive galaxies by triggering both intense starbursts and rapid black hole growth. This process may have been followed by active galactic nuclei (AGN) feedback and/or gas stripping which likely quenched star formation, leading to the formation of massive quiescent galaxies. To understand this evolutionary path, it is crucial to clarify how AGN feedback and interactions affect molecular gas and whether they enhance and/or suppress star formation \citep[e.g.,][]{Hopkins_2008ApJS,Poggianti_2017ApJ,Spilker_2022ApJ}. Although high spatial resolution observations of molecular gas in closely interacting systems at high-\textit{z} remain challenging, compact galaxy groups (CGGs) in the local universe serve as valuable analogs. 
These groups, first characterized by \citet{Hickson1982ApJ}, consist of a small number of galaxies in close proximity and are known to undergo frequent interactions and mergers. They are expected to evolve into massive elliptical galaxies \citep[e.g.,][]{Barnes1989}, and thus provide key laboratories for studying the dynamical and environmental processes that regulate star formation and gas evolution in dense environments.

CGGs often contain large reservoirs of intergalactic gas and show signs of intergalactic stripping and tidal interactions. Such processes can both suppress and enhance star formation. While gas stripping can remove star-forming material from galaxies, potentially leading to quenching, the mixing of stripped interstellar material with the intergalactic medium (IGM) can promote cooling and accretion, triggering new star formation activity in the IGM \citep[e.g.,][]{Lisenfeld_2002AA}. Understanding the relative importance of these processes remains an open question. Studying CGGs offers a unique opportunity to explore how galaxy interactions affect molecular gas and star formation. Importantly, this is not only relevant for understanding the early universe. About half of the galaxies in the local universe ($z\sim0.01$) are members of galaxy groups \citep[e.g.,][]{Makarov_2011MN}. Thus, CGGs also play a critical role in shaping galaxy evolution at low redshift.

\begin{figure*}[t!]
 \begin{center}
  \includegraphics[width=180mm]{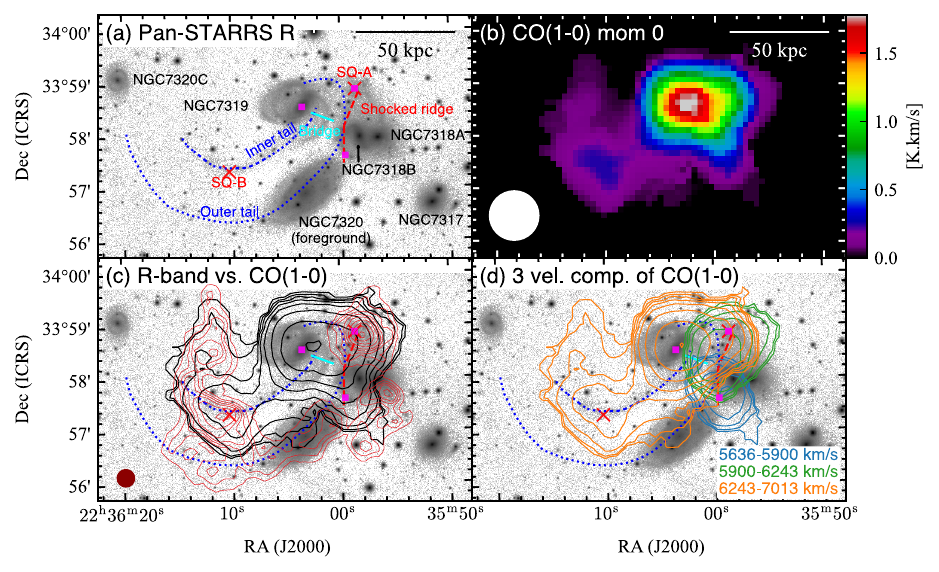} 
 \end{center}
\caption{
(a) Pan-STARRS $R$-band image of SQ with labels indicating the names of host galaxies and key structures. 
The figure design is inspired by  \citep{Renaud_2010ApJ}. 
(b) CO(1-0) moment 0 map of SQ. The white filled circle in the lower left corner represents the beam size of the TP. The TP field of view ($7^{\prime}.4 \times 7^{\prime}.0$) is larger than the displayed area, but no emission was detected outside it.
(c) CO(1-0) moment 0 map superposed on the $R$-band image, with black contours at $\log I_\mathrm{CO(1-0)}/(\mathrm{K~km~s^{-1}})=-1.50,-1.25,-1.00,-0.75,-0.50,-0.25,0.00,0.25$.
Red contours represent the HI moment 0 map from \citet{Williams_2002AJ}; contour levels are 5.8, 15, 23, 32, 44, 61, 87, 120, and 180~$\times~10^{19}~\mathrm{cm^{-2}}$, with a beam size of $19^{\prime\prime}.4\times18^{\prime\prime}6$.
(d) Same as panel (c), but showing the moment-0 maps separated into three velocity components (see Section~3.2).
The magenta squares indicate the coordinates of the profiles shown in Figure~\ref{fig:SQprofile}.}\label{fig:SQmom0}
\end{figure*}

Stephan’s Quintet (SQ; Figure~\ref{fig:SQmom0}(a)) is a well-known compact galaxy group located at a distance of 88.6 Mpc (from the NASA/IPAC Extragalactic Database), which is composed of five galaxies: NGC 7317, 7318A, 7318B, 7319, and 7320C (with NGC 7320 being a foreground galaxy). SQ is in an advanced stage of dynamical interaction, as indicated by multiple tidal tails that are clearly visible in deep optical imaging \citep[e.g.,][]{Fedotov2011AJ, Duc_2018MNRAS}, with ages of 150–200 Myr (inner tail) and 400–500 Myr (outer tail) created by successive encounters \citep{Fedotov2011AJ}.
A prominent large-scale shocked ridge (red dashed line in Figure~\ref{fig:SQmom0}(a)), clearly visible in the infrared \citep[e.g.,][]{Pontoppidan_2022ApJ}, is produced by the ongoing collision between the intruder galaxy NGC 7318B and both NGC 7319 and the tidal debris from previous interactions \citep{Allen_Hartsuiker_1972,Ohyama_1998ApJ}. This collision occurs as NGC 7318B enters the group from behind at a relative line-of-sight velocity of $\sim 900~\mathrm{km~s^{-1}}$ \citep{Xu_2003ApJ}. NGC 7319, classified as a Seyfert 2 galaxy with a large-scale outflow \citep{Aoki_1996AJ}, is thought to have been involved in nearly all past interactions. As a result of these repeated encounters, most of HI gas has been stripped from the galaxies and now resides in the IGM \citep{Williams_2002AJ}. Remarkably, new intergalactic star-forming regions, such as SQ-A and SQ-B (red crosses in Figure~\ref{fig:SQmom0}(a)), have emerged even within the stripped HI gas structures \citep[e.g.,][]{Lisenfeld_2002AA}, highlighting the complex interplay between gas removal and induced star formation in the group environment.

In this paper, we present the results of CO(1–0) mapping of SQ using Atacama Large Millimeter/submillimeter Array (ALMA) Total Power (TP), achieving a uniform noise level across the mapped area and aiming to reveal the total CO flux and its spatial distribution across the entire system. Numerous CO observations of SQ have been conducted \citep[e.g.,][]{Yun_1997ApJ,Gao_Xu_2000ApJ,Smith_2001AJ,Lisenfeld_2002AA,Lisenfeld_2004AA,Petitpas_2005ApJ,Guillard_2012ApJ,Yttergren_2021AA,Appleton_2023ApJ,Emonts_2025ApJ}. However, previous CO observations of SQ have mainly focused on the main member galaxies (NGC~7318AB and NGC~7319), the shocked ridge, and SQ-A. The tidal tail has not been fully covered, with observations limited to the SQ-B region.
Moreover, most mappings targeted only limited areas and/or relied solely on interferometric data, which often suffer from significant missing flux. Therefore, the total amount of molecular gas across the entire system has remained unknown. While the TP data have lower spatial resolution, they are free from uncertainties that can potentially be introduced in the imaging process of interferometric data, making them highly valuable for investigating the global properties of molecular gas in the system.

\begin{figure*}[t!]
 \begin{center}
  \includegraphics[width=180mm]{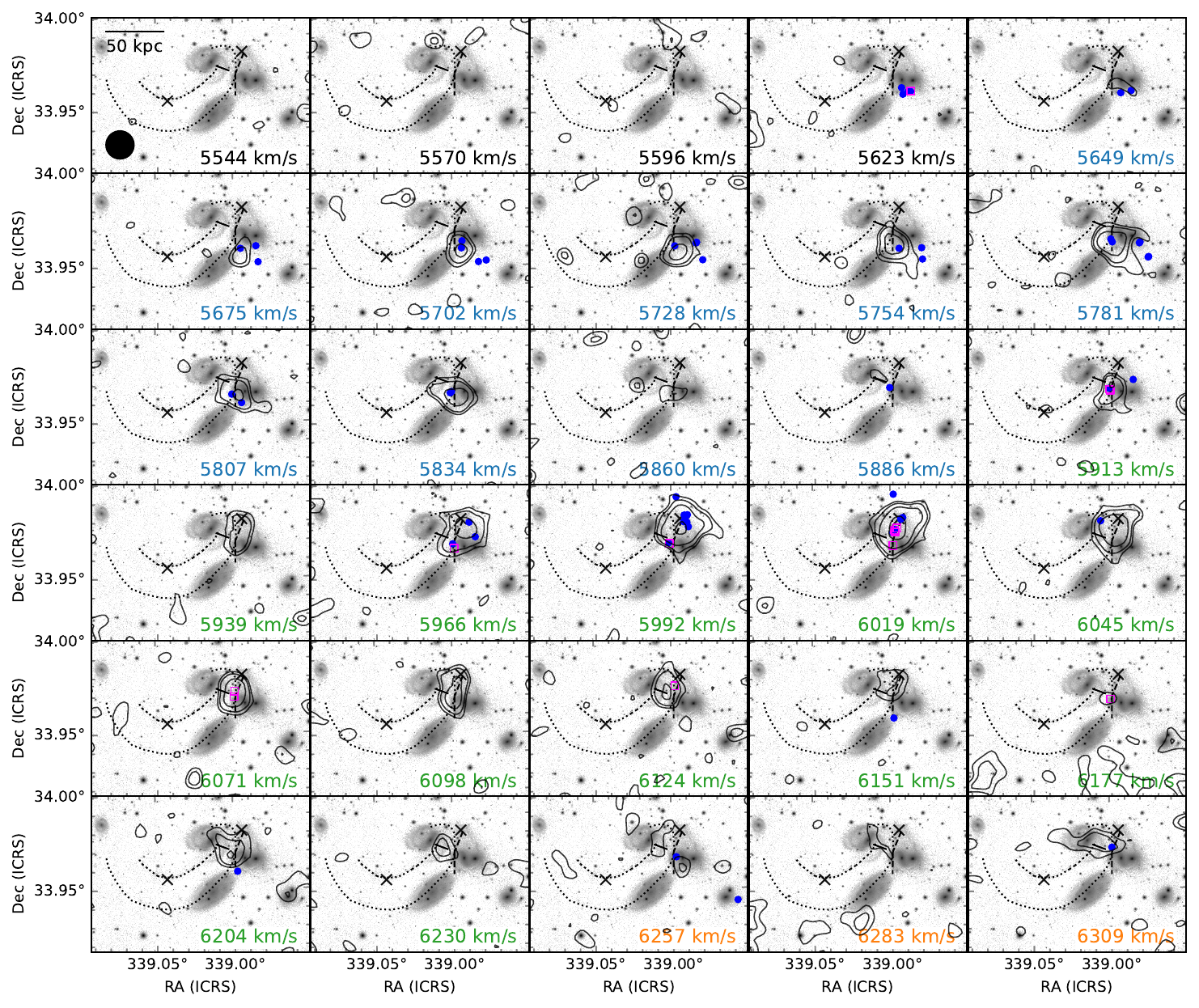} 
 \end{center}
\caption{
Channel maps of the CO(1--0) of SQ superposed on the same $R$-band image as in Figure~\ref{fig:SQmom0}. The contours are $2\sigma_\mathrm{rms}$, $3\sigma_\mathrm{rms}$, $5\sigma_\mathrm{rms}$, $10\sigma_\mathrm{rms}$, $15\sigma_\mathrm{rms}$, and $20\sigma_\mathrm{rms}$.  
The velocity labels are color-coded to correspond to the three velocity components shown in Figure~\ref{fig:SQmom0}(d).
All symbols indicate the positions of H$\alpha$ emitters in position–position-velocity space identified by \citet{Duarte_Puertas_2021AA}.  
Blue-filled circles represent emitters classified as HII regions based on the BPT diagram.
Magenta open squares represent composite based on the BPT diagram. Velocities are given in optical velocity. The crosses and dashed guide lines are the same as in Figure~\ref{fig:SQmom0}.}\label{fig:SQchmap1}
\end{figure*}

\begin{figure*}[t!]\ContinuedFloat
 \begin{center}
  \includegraphics[width=180mm]{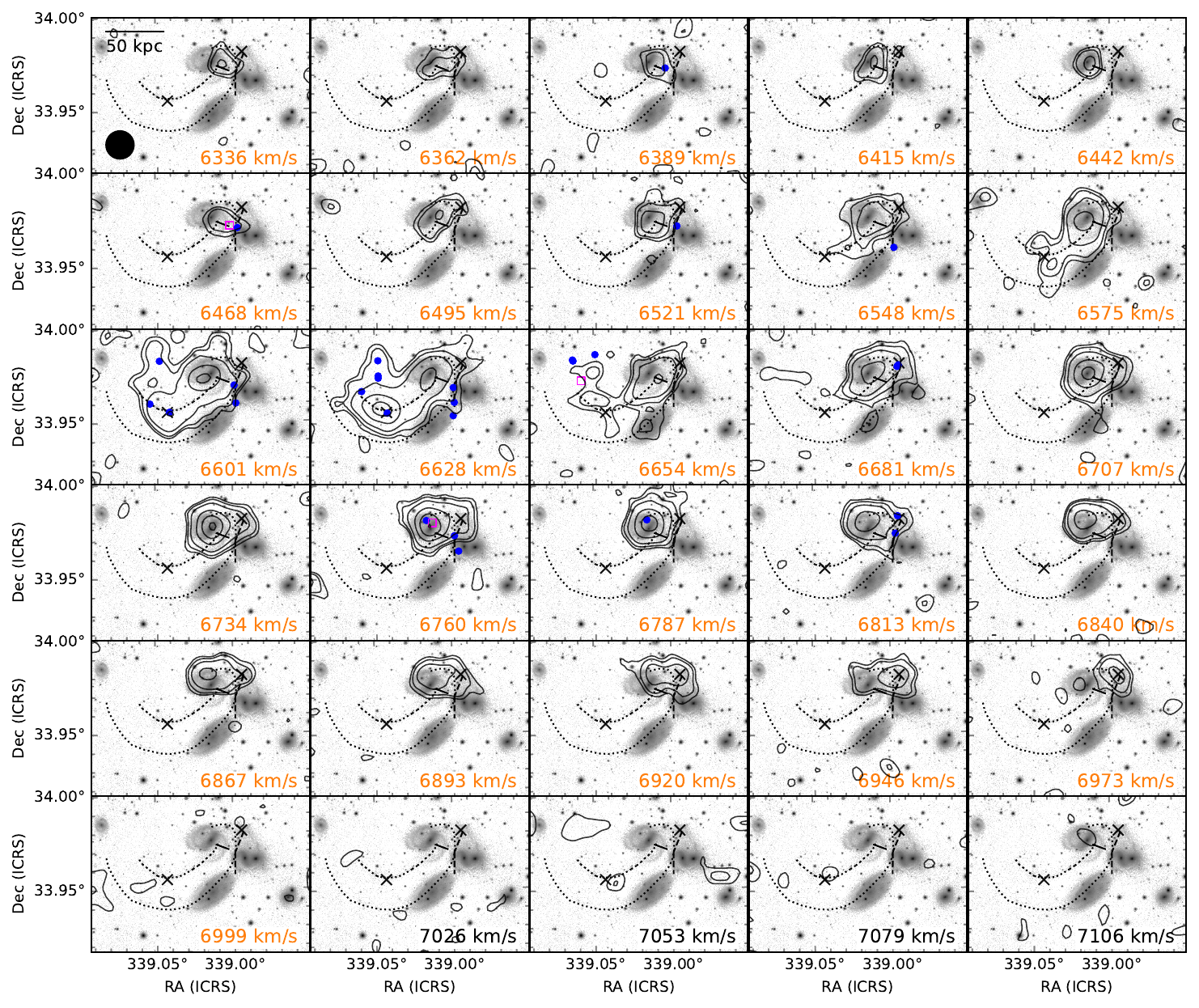} 
 \end{center}
\caption{
Continued. }\label{fig:SQchmap2}
\end{figure*}

\section{Observation and Data Reduction}

The CO(1–0) mapping of the entire SQ was conducted with ALMA TP as part of project 2023.1.01101.S (PI: F. Maeda). The field of view (FoV) spans $7^{\prime}.4 \times 7^{\prime}.0$, centered at ($22^\mathrm{h}36^\mathrm{m}05^\mathrm{s}.11$, $33^\circ57^\prime58^{\prime\prime}.1$),  covering all previously observed HI emission \citep{Williams_2002AJ} and HII regions \citep{Duarte_Puertas_2019AA} in the system. On-the-fly observations were carried out between October 2, 2023, and May 4, 2024, resulting in a total of 197 execution blocks (EBs). The average number of TP antennas per EB was 2.72. The total observing time was 197.2 hours, with an on-source integration time of 107.0 hours.  The ACA spectrometer was used to obtain spectra with a bandwidth of 2.000~GHz and a channel spacing of 976.562~kHz, corresponding to $\sim2.5~\mathrm{km~s^{-1}}$. The spectral window was centered at 112.697~GHz, corresponding to the redshifted frequency of the $\mathrm{CO(1-0)}$ line at an optical velocity of 6861~$\mathrm{km~s^{-1}}$.
The median system temperature for each EB typically ranged from 80 to 120~K.

Calibration was performed using CASA (version 6.5.4.9) with the observatory-provided script to reproduce the standard pipeline processing (Pipeline version 2023.1.0.124). However, during baseline subtraction, the pipeline-defined fitting range did not adequately mask the CO line. Therefore, we modified the masking range to exclude the frequency range from 112.578~GHz to 113.172~GHz, which corresponds to an optical velocity range of 5550–7150~$\mathrm{km~s^{-1}}$, where CO emission was detected in SQ. 
Each EB was imaged using CASA task \verb|sdimaging|, and the resulting cubes were combined using the weights provided by \verb|sdimaging|. Baseline subtraction was successful in regions like NGC~7319, the shocked ridge, and SQ-A, where emission spanned across the masked velocity range. However, in SQ-B and the surrounding tidal tail, where the CO line width is much lower ($\leq 120~\mathrm{km~s^{-1}}$), residual baseline features remained. 
To mitigate these residuals, we applied a high-pass filter using fast Fourier transform (FFT) specifically to spatial pixels within these low-linewidth regions. 
Prior to filtering, the velocity ranges with CO emission were masked. The spectra were then converted to the frequency domain, low-frequency components below $1/1000~\mathrm{(km~s^{-1})^{-1}}$ were removed, and the data were transformed back into the velocity domain.
The final cube, binned to 25~$\mathrm{km~s^{-1}}$ resolution, has a beam size of $58^{\prime\prime}.0$ corresponding to 24.9~kpc, and a uniform rms noise level of $\sigma_\mathrm{rms}=0.31$~mK.

\section{Results and Discussions}

\subsection{Global properties}

\begin{figure}[t!]
 \begin{center}
  \includegraphics[width=85mm]{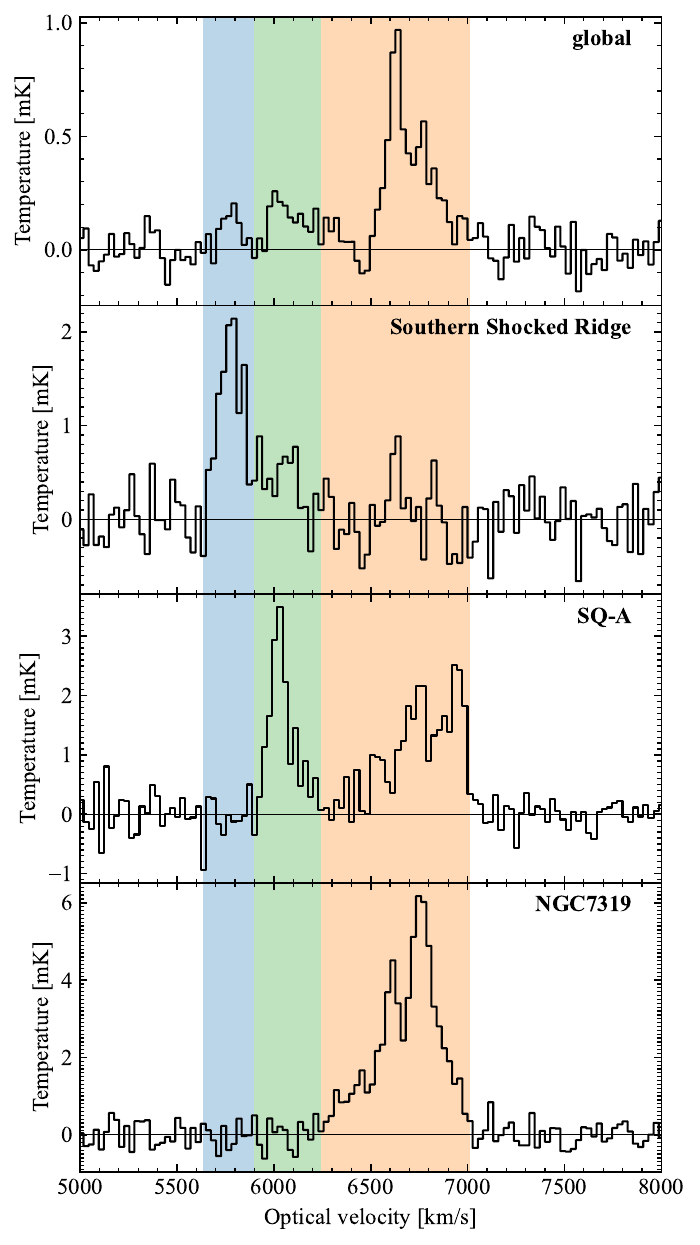} 
 \end{center}
\caption{
Global CO(1-0) profile (top panel) and representative CO(1–0) profiles at selected positions (lower panels). The global profile is extracted from the rectangular region that covers all of the CO(1–0) emission.
The selected positions are indicated in Figure~\ref{fig:SQmom0}.
The blue, green, and orange bands correspond to the three velocity components.}\label{fig:SQprofile}
\end{figure}

The resultant CO(1–0) integrated intensity (moment-0) maps are shown in Figure~\ref{fig:SQmom0}(b–d). Significant emission was identified within the data cube by selecting voxels with signal exceeding 5.0$\sigma_\mathrm{rms}$ in at least two consecutive velocity channels. These regions were then expanded to include all adjacent pixels with signal above 2.0$\sigma_\mathrm{rms}$.

We find that the molecular gas in SQ is distributed over a wide area spanning $\sim120\times80~\mathrm{kpc}$, not only around NGC~7319, the shocked ridge, and SQ-A, but also broadly across the inner tail and its northern region (Figures~\ref{fig:SQmom0}(b-c)).
The extended molecular gas is different from the HI map by \citet{Williams_2002AJ} shown as Figure~\ref{fig:SQmom0}(c). Strong CO emission is detected in NGC~7319, despite its HI gas being stripped\footnote{The VLA observations by \citet{Williams_2002AJ}  resolved structures smaller than $15^\prime$, so the non-detection of HI in NGC~7319 implies a 3$\sigma$ upper limit on the column density of $\leq 5 \times 10^{19}~\mathrm{cm^{-2}}$.}, indicating the presence of remaining molecular gas. CO is also seen along the shocked ridge. A new finding from our observations is that molecular gas extends from NGC~7319 along the inner tail and further north beyond the tail. The strongest CO emission within the tail corresponds to SQ-B. We identified three main velocity components in SQ (Figure~\ref{fig:SQmom0}(d)), which will be discussed in the next subsection.

The total CO(1--0) luminosity is $L^\prime_\mathrm{CO(1-0)}
=(2.47\pm0.12)\times10^9~\mathrm{K~km~s^{-1}~pc^2}$, accounting for the absolute flux calibration accuracy of $\pm 5\%$ (ALMA Technical Handbook)\footnote{https://almascience.nao.ac.jp/documents-and-tools/cycle10/alma-technical-handbook}. 
Assuming the Galactic CO-to-H$_2$ conversion factor of $\alpha_\mathrm{CO}=4.35~M_\odot~\mathrm{ (K~km~s^{-1}~pc^2)^{-1}}$ \citep{bolatto_conversion_2013}, the total molecular gas mass is estimated to be $(1.07\pm0.05)\times 10^{10}~M_\odot$. We adopt this conversion factor because most HII regions in SQ are metal-rich, with $12+\log(\mathrm{O/H})\geq8.5$ based on the O3N2 method \citep[][]{Duarte_Puertas_2021AA}. However, $\alpha_\mathrm{CO}$ is uncertain and likely varies depending on the environment. For example, in the shock region, the large velocity width can decrease the optical depth of the CO line, resulting in a lower $\alpha_\mathrm{CO}$ than the Galactic value, as reported in other interacting galaxies \citep{Braine_2003,Zhu_2007AJ}. 
Therefore, the total molecular gas mass assuming the Galactic $\alpha_{\rm CO}$ may represent an upper limit.
Addressing these uncertainties will require higher-resolution observations and measurements of additional CO transitions and isotopologues, which we leave for future work.

The star formation rate (SFR) of SQ over the whole FoV in Figure~\ref{fig:SQmom0} is estimated to be 3.1~$M_\odot~\mathrm{yr^{-1}}$ from dust-corrected spectroscopic H$\alpha$ emission confirmed as star-formation origin via the Baldwin, Philips, \& Terlevich (BPT) diagram of $\mathrm{[O III]/H\beta}$ versus $\mathrm{[NII]/H\alpha} $ \citep{Duarte_Puertas_2021AA}, and 7.5~$M_\odot~\mathrm{yr^{-1}}$ from dust-corrected UV emission \citep{Natale_2010ApJ}. These values yield a global star formation efficiency (SFE) for SQ of 0.29–0.70~$\mathrm{Gyr^{-1}}$, defined as the ratio of SFR to molecular gas mass.
This value is comparable to or lower than that of nearby star-forming galaxies \citep[$0.5–1.0~\mathrm{ Gyr^{-1}}$][]{Leroy2008AJ,Muraoka_2019PASJ,Sun_2023ApJ}, indicating that star formation is not globally efficient in the system. It is also consistent with previous survey results of CGG member galaxies \citep[median value of $0.38~\mathrm{Gyr^{-1}}$][]{Alatalo_2015ApJ}, suggesting that violent interactions tend to suppress star formation across the system. 

Here, we compare our results with recent observations. \citet{Emonts_2025ApJ} observed SQ (excluding the tail) in CO(2–1) using only the ALMA 7m array. They assumed a CO(2–1)/CO(1–0) ratio ($R_{21}$) of 0.51 and, based on this, estimated a $L^\prime_\mathrm{CO(1-0)}$ of $1.24\times10^9~\mathrm{K~km~s^{-1}~pc^2}$. In the same region, our moment-0 map yields $2.20\times10^9~\mathrm{K~km~s^{-1}~pc^2}$, which is  1.8 times higher.
This difference may arise from CO(2–1) emission extended on scales larger than the maximum recoverable scale of  7m array ($\sim$12 kpc) and/or from an inaccurate $R_{21}$. The adopted $R_{21}$ of 0.51 was derived by comparing 7m CO(2–1) with single-pointing Combined Array for Research in Millimeter-wave Astronomy (CARMA) CO(1–0) data in part of SQ, both of which likely suffer from missing flux. Verifying the presence of extended CO(2–1) emission and accurately measuring $R_{21}$ and its spatial distribution will require future single-dish (e.g., TP) CO(2–1) observations.


\subsection{Velocity structures}
\label{sec: velocity structures}
Figure~\ref{fig:SQchmap1} and the top panel of Figure~\ref{fig:SQprofile} present the CO(1–0) channel maps and the global CO(1-0) profile, respectively, revealing molecular gas detected over a velocity range of $\sim1300~\mathrm{km~s^{-1}}$, from approximately 5650 to 7000~$\mathrm{km~s^{-1}}$ in optical velocity. 
Multiple kinematically distinct components are seen, which is also reported in previous CO observations \citep{Lisenfeld_2002AA,Guillard_2012ApJ,Emonts_2025ApJ}.

From the global profile (top panel of Figure~\ref{fig:SQprofile}), we can identify three distinct velocity components. We define these as 5636–5900 $\mathrm{km~s^{-1}}$ (low-velocity component), 5900–6243 $\mathrm{km~s^{-1}}$ (mid-velocity component), and 6243–7013 $\mathrm{km~s^{-1}}$ (high-velocity component). These boundaries were determined as follows. Based on the channel maps (Figure~\ref{fig:SQchmap1}), the component associated with the southern shocked ridge and NGC~7318B drops sharply just below 5900~$\mathrm{km~s^{-1}}$, while the component associated with SQ-A becomes prominent from around 5900~$\mathrm{km~s^{-1}}$. We therefore set the low–mid boundary to 5900~$\mathrm{km~s^{-1}}$. The mid-velocity component associated with the SQ-A component extends up to $\sim$6200~$\mathrm{km~s^{-1}}$, above which the NGC~7319 component emerges; this is adopted as the mid–high velocity boundary. To illustrate these transitions more clearly, we show CO(1–0) spectra in the southern shocked ridge, SQ-A, and the center of NGC 7319 in Figure~\ref{fig:SQprofile}.
Note that our division of the velocity components differs from that of \citet{Emonts_2025ApJ}, which may be due to differences in the observed transition (CO(1–0) versus CO(2–1)) and the observing instruments (TP versus 7m array).

\begin{figure*}[t!]
 \begin{center}
  \includegraphics[width=180mm]{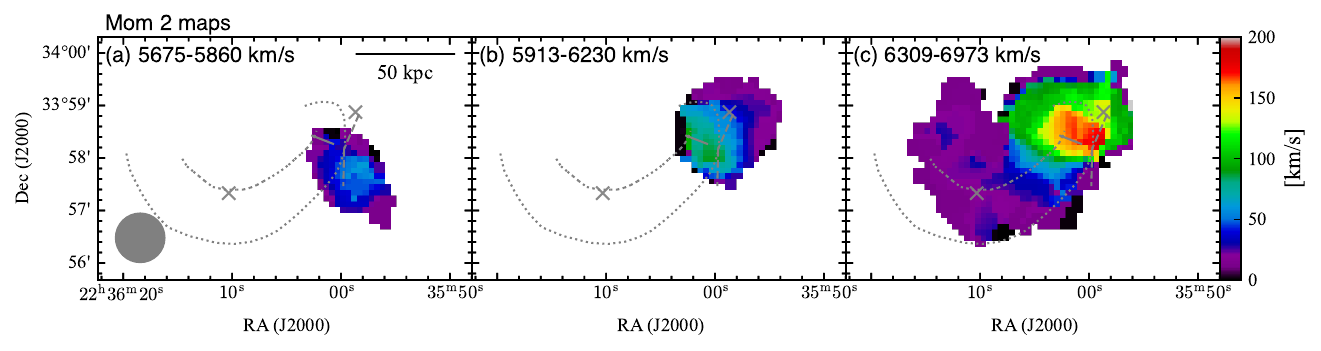} 
 \end{center}
\caption{
CO(1–0) moment 2 maps for each of the three velocity components in SQ.}\label{fig:SQmom2}
\end{figure*}

The moment-0 and velocity dispersion (moment-2) maps for each component are shown in Figures~\ref{fig:SQmom0}(d) and \ref{fig:SQmom2}, respectively.
Figure~\ref{fig:SQchmap1} also marks the positions of identified HII regions by \citet{Duarte_Puertas_2021AA}.
In the following, we examine each velocity component in detail, considering its kinematic in relation to star formation activity.

{\bf Low velocity component (5636–5900~$\mathrm{km~s^{-1}}$):} 
This component is distributed along the shocked ridge and is closely associated with NGC~7318B, whose systemic velocity is 5765~$\mathrm{km~s^{-1}}$ \citep{Sulentic_2001AJ}. 
The velocity dispersion is relatively low, around 50~$\mathrm{km~s^{-1}}$ (Figure~\ref{fig:SQmom2}(a)), and the line-of-sight velocities of molecular gas and HII regions are generally consistent along the shocked ridge, indicating that active star formation is ongoing. The SFR derived from extinction corrected H$\alpha$ in this velocity range is 0.92~$M_\odot~\mathrm{yr^{-1}}$ \citep{Duarte_Puertas_2021AA}, the molecular gas mass is $(6.78\pm0.34)\times10^8~M_\odot$, and the SFE is 1.36~$\mathrm{Gyr^{-1}}$. 

{\bf Mid velocity component (5900–6243~$\mathrm{km~s^{-1}}$):}
This component connects to the low-velocity component, extending from the north side of the shocked ridge to SQ-A. 
Optical emission-line data indicate shock influence; \citet{Duarte_Puertas_2021AA} showed that H$\alpha$ emitters here appear as composites on the BPT diagram (magenta squares in Figure~\ref{fig:SQchmap1}) are consistent with a fast shock model
\citep[shock velocity of 175–1000~$\mathrm{km~s^{-1}}$][]{Allen_2008ApJS}. Despite this, CO velocity dispersions are 50–80~$\mathrm{km~s^{-1}}$ along the shocked ridge but $<50~\mathrm{km~s^{-1}}$ in SQ-A, where most HII regions are concentrated. 
The total SFR, molecular gas mass, and SFE of this component are 0.62~$M_\odot~\mathrm{yr^{-1}}$, $(1.62\pm0.08)\times10^9~M_\odot$, and 0.38~$\mathrm{Gyr^{-1}}$, respectively.

{\bf High velocity component (6243–7013~$\mathrm{km~s^{-1}}$):}
This component is associated with NGC~7319, whose systemic velocity is 6550~$\mathrm{km~s^{-1}}$ \citep{Sulentic_2001AJ}. 
The total SFR, molecular gas mass, and SFE are 1.62~$M_\odot~\mathrm{yr^{-1}}$, $(8.43\pm0.42)\times 10^9~M_\odot$, and 0.19~$\mathrm{Gyr^{-1}}$, respectively. A key feature revealed by our observations is the molecular gas extending from NGC~7319 along the inner tail (6550–6650~$\mathrm{km~s^{-1}}$).
The tail exhibits a small velocity dispersion ($\sim$20~$\mathrm{km~s^{-1}}$), whereas the other regions (NGC~7319 and SQ-A) show higher values of $50-200~\mathrm{km~s^{-1}}$ (Figure~\ref{fig:SQmom2}(c)). The tail component is discussed further in the next subsection.
Excluding the tail, the molecular gas in the high velocity component distributes around NGC~7319 and SQ-A, totaling $(6.79 \pm 0.34) \times 10^9~M_\odot$ (63\% of the system).
Here, we summed the molecular gas mass in a region with velocity dispersions above 50 $\mathrm{km~s^{-1}}$ in Figure~\ref{fig:SQmom2}(c).
The HII regions associated with these relatively high velocity dispersion components have a combined SFR of 1.10~$M_\odot~\mathrm{yr^{-1}}$ and an SFE of 0.16~$\mathrm{Gyr^{-1}}$, indicating that star formation is generally suppressed. Notably, two HII regions in SQ-A, at 6681~$\mathrm{km~s^{-1}}$ (see Figure~\ref{fig:SQchmap1}), account for a combined SFR of 0.87~$M_\odot~\mathrm{yr^{-1}}$, suggesting that the molecular gas directly associated with NGC~7319 exhibits significantly suppressed star formation activity. \citet{Emonts_2025ApJ} argued that this suppression is likely caused by the influence of NGC~7319’s central AGN, specifically through jet-driven outflows or turbulence resulting from interactions with the radio jet.

The molecular gas in SQ-A has been thought to originate from the compression of gas located in the outskirts of galaxies or from diffuse intergalactic gas during the collision
\citep{Lisenfeld_2002AA, Iglesias_2012AA}. However, our observations newly reveal that in the 6681–6946~$\mathrm{km~s^{-1}}$ channel map, molecular gas appears continuously connected between NGC~7319 and SQ-A, suggesting an additional contribution from gas stripped from NGC~7319.

{\bf Bridge:} A structure extending perpendicular to the shocked ridge, between NGC~7319 and the ridge, is known to bridge the gap between the low-/mid-velocity and high-velocity components 
\citep{Guillard_2012ApJ,Emonts_2025ApJ}.
Our observations confirm this feature at 6100–6400~$\mathrm{km~s^{-1}}$, as shown in Figure~\ref{fig:SQchmap1}, consistent with previous single-dish observations \citep{Guillard_2012ApJ}. It connects the low-/mid-velocity and high-velocity components, exhibiting very broad line widths indicative of highly turbulent gas.

\subsection{Molecular gas in tidal tails}
\label{sec: tail}

Our wide-field mapping reveals that molecular gas extends across the inner tail of SQ and its northern region, spanning a projected scale of  $\sim$100~kpc. A key feature is the very low velocity width, with a velocity dispersion of about 20~$\mathrm{km~s^{-1}}$ (Figure~\ref{fig:SQmom2}(c)). As seen in Figure~\ref{fig:SQchmap1} , the CO and H$\alpha$ line-of-sight velocities align well within this low velocity range (indicate velocity range, i.e., 6601-6654~$\mathrm{km~s^{-1}}$), suggesting ongoing star formation in these regions.
The molecular gas mass in the tail is $(1.64\pm0.08)\times10^9~M_\odot$, accounting for 15\% of the total molecular gas mass of SQ. Here, we summed the molecular gas mass in a region east of NGC~7319 with velocity dispersions below 50 $\mathrm{km~s^{-1}}$ in Figure~\ref{fig:SQmom2}(c).
With a total SFR of 0.52~$M_\odot~\mathrm{yr^{-1}}$, the corresponding SFE is 0.32~$\mathrm{Gyr^{-1}}$. 

Previous HI observations \citep{Williams_2002AJ} showed that HI gas was stripped from NGC~7319, extends along the outer tail, deviates from it partway, passes through SQ-B, and continues northward. In contrast, our CO observations reveal a different distribution (Figure~\ref{fig:SQmom0}(c)): the molecular gas extends along the inner tail up to SQ-B and only beyond SQ-B follows the same northern extension seen in HI.

Molecular gas in tidal tails has been studied in various nearby interacting galaxies \citep{braine_formation_2000,braine_abundant_2001,Lisenfeld_2002AA,Lisenfeld_2004AA,lisenfeld_molecular_2008,lisenfeld_molecular_2016,querejeta_TDG_2021,Spilker_2022ApJ,Kovakkuni_2023,Maeda_2024}. Two main scenarios of origin of the molecular gas have been proposed: in situ formation from HI \citep[e.g.,][]{braine_formation_2000} and stripping from parent galaxies \citep[e.g.,][]{Spilker_2022ApJ}. For SQ-B, previous observations have favored the in situ scenario. \citet{Lisenfeld_2002AA}, using IRAM 30m observations, found that the molecular and HI gas velocities in SQ-B match each other, suggesting the molecular gas formed locally by conversion from the tidal HI gas under local pressure or shocks. 
The low metallicity observed across the tail generally supports the idea that this gas originated from HI stripped during the collision ~$\sim$400~Myr ago that formed the outer tail \citep{Fedotov2011AJ}.
However, our new observations suggest an additional contribution: unlike HI, molecular gas traces the inner tail from NGC~7319 toward SQ-B (Figure~\ref{fig:SQmom0}(c)), implying that some of the molecular gas in SQ-B may have been stripped from NGC~7319 during the more recent 150–200~Myr-old collision that formed the inner tail  \citep{Fedotov2011AJ}. 
Therefore, both the in situ formation and stripping scenarios seem to coexist, as seen in SQ-A.

The content discussed in Sections \ref{sec: velocity structures} and \ref{sec: tail} is based on the TP data alone. However, the limited resolution of the TP observations may partly cause the apparent connection between NGC~7319 and regions such as SQ-A or SQ-B, as the large beam size can blend molecular gas from different regions. To obtain a more detailed and conclusive picture of the spatial structure and more rigorously test the stripping scenario, we plan to combine the TP data with CO(1–0) observations from the ALMA 7m array, which will be presented in our next paper (Yamamoto et al., in prep.).

\section{Conclusions}

We conducted the first wide-field ALMA TP CO(1–0) mapping of Stephan’s Quintet with a uniform noise level at a spatial scale of $\sim25~\mathrm{kpc}$, revealing the spatially and dynamically extended molecular gas component across the system. Our key findings are:
\begin{enumerate}
    \item Molecular gas is detected over $\sim120\times80~\mathrm{kpc}$, spanning the NGC~7319, NGC~7318B, the shocked ridge, SQ-A, and notably the inner tidal tail and the intergalactic regions north of the inner tail. The extended molecular gas is inconsistent with HI morphology.
    \item The total CO(1–0) luminosity is  $L^\prime_\mathrm{CO(1-0)}=(2.47\pm0.12) \times10^9~\mathrm{K~km~s^{-1}~pc^2}$, corresponding to a molecular gas mass of $(1.07 \pm 0.05)\times10^{10}~M_\odot$ assuming the Galactic $\alpha_\mathrm{CO}$. The global SFE of the system is estimated to be $0.29-0.70~\mathrm{Gyr^{-1}}$, which is comparable to or lower than that of nearby star-forming galaxies.
    \item Our CO(1–0) mapping reveals molecular gas spread across a velocity range of $\sim1300~\mathrm{km~s^{-1}}$, which can be broadly divided into three main components (low, mid, high), each containing both turbulent and settled gas that locally enhances or suppresses star formation. The low- and mid-velocity components, linked to NGC~7318B and the shocked ridge, show active star formation, while the high-velocity component, associated with NGC~7319, shows suppressed star formation. In SQ-A, both in situ molecular gas formation and stripping from NGC~7319 likely coexist, as suggested by the combination of low- and high-metallicity HII regions, and supported by our discovery of molecular gas connecting from NGC~7319.
    \item Our mapping shows that molecular gas extends $\sim$100~kpc along the inner tail and north of it, containing $(1.64\pm0.08)\times10^9~M_\odot$ (15\% of the total) with low velocity dispersions ($\sim20~\mathrm{km~s^{-1}}$) and ongoing star formation. While previous studies suggested in situ formation for SQ-B, our data indicate a possible additional contribution from gas stripped from NGC~7319.
\end{enumerate}

\begin{acknowledgments}
We would like to thank the referee for useful comments.
We are deeply grateful to the staff of the Nobeyama Radio Observatory for granting us the observing time for the pilot CO(1–0) observations of the tidal tail in Stephan’s Quintet, which formed the basis for our subsequent ALMA project.
F.M. is supported by JSPS KAKENHI grant No. JP23K13142.  S.K. is supported by JSPS KAKENHI grant No. JP25K07371.
K.O. is supported by JSPS KAKENHI grant No. JP23K03458.
A.K. is supported by JSPS KAKENHI grant No. JP24KL1904.
K.K. acknowledges the support by JSPS KAKENHI Grant Numbers JP22H04939, JP23K20035, and JP24H00004. 
M.I.N.K. is supported by JSPS KAKENHI grant No. JP22K14080.
T.T. acknowledges the support by JSPS KAKENHI Grant No. 20H00172. 
A.T. is supported by JSPS KAKENHI grant No. 24KJ0562.
This paper makes use of the following ALMA data:
ADS/JAO.ALMA \#2023.1.01101.S.
ALMA is a partnership of ESO (representing its member states), NSF (USA), and NINS (Japan), together with NRC (Canada), MOST and ASIAA (Taiwan), and KASI (Republic of Korea), in cooperation with the Republic of Chile. The Joint ALMA Observatory is operated by ESO, AUI/NRAO, and NAOJ. 
Data analysis was in part carried out on the Multi-wavelength Data Analysis System operated by the Astronomy Data Center (ADC), NAOJ.
This research also made use of APLpy, an open-source plotting package for Python \citep{Robitaille_2012}.
\end{acknowledgments}





%
\facilities{ALMA}

\software{CASA \citep{CASA_2022PASP}, Astropy \citep{astropy_2018}, APLpy \citep{Robitaille_2012}}



\bibliography{SQ}{}
\bibliographystyle{aasjournalv7}



\end{document}